# ANALYSIS OF MANAGERIAL BEHAVIORS IN BUSINESS MANAGEMENT

Ernest GÓRKA[1*], Dariusz BARAN[2], Michał ĆWIĄKAŁA[3], Gabriela WOJAK[4],
Robert MARCZUK[5], Katarzyna OLSZYŃSKA[6], Piotr MRZYGŁÓD[7],
Maciej FRASUNKIEWICZ[8], Piotr RĘCZAJSKI[9], Kamil SAŁUGA[10],
Maciej ŚLUSARCZYK[11], Jan PIWNIK[12]

[1] Wyższa Szkoła Kształcenia Zawodowego; ernest.gorka@wskz.pl, ORCID: 0009-0006-3293-5670
[2] Pomorska Szkoła Wyższa w Starogardzie Gdańskim, Instytut Zarządzania, Ekonomii i Logistyki;
dariusz.baran@twojestudia.pl, ORCID: 0009-0006-8697-5459
[3] Wyższa Szkoła Kształcenia Zawodowego; michal.cwiakala@wskz.pl, ORCID: 0000-0001-9706-864X
[4] I'M Brand Institute sp. z o.o.; g.wojak@imbrandinstitute.com, ORCID: 0009-0003-2958-365X
[5] I'M Brand Institute sp. z o.o.; r.marczuk@imbrandinstitute.com, ORCID: 0009-0008-3553-6581
[6] Polsko Japońska Akademia Technik Komputerowych; kontakt@olszynska.com,
ORCID: 0009-0003-4309-6233
[7] Piotr Mrzygłód Sprzedaż-Marketing-Consulting; piotr@marketing-sprzedaz.pl, ORCID: 0009-0006-5269-0359
[8] F3-TFS sp. z o.o.; m.frasunkiewicz@imbrandinstitute.com, ORCID: 0009-0006-6079-4924
[9] MAMASTUDIO Pawlik, Ręczajski, sp. j.; piotr@mamastudio.pl, ORCID: 0009-0000-4745-5940
[10] I'M Brand Institute sp. z o.o.; k.saluga@imbrandinstitute.com, ORCID: 0009-0000-0440-4035
[11] Pomorska Szkoła Wyższa w Starogardzie Gdańskim, Instytut Zarządzania, Ekonomii i Logistyki;
maciej.slusarczyk@twojestudia.pl, ORCID: 0000-0001-6612-8179
[12] WSB Merito University in Gdańsk, Faculty of Computer Science and New Technologies;
jpiwnik@wsb.gda.pl, ORCID: 0000-0001-9436-7142
* Correspondence author

**Purpose:** The primary aim of this research is to empirically investigate the situational adaptability of managerial behaviors and their impact on organizational effectiveness through the practical application of Kenneth Blanchard's diagnostic tool.

**Design/methodology/approach**: The objectives are achieved by employing a scenario-based diagnostic questionnaire derived from Kenneth Blanchard's situational leadership model to analyze managerial behaviors and their adaptability in ten selected companies.

**Findings:** The research revealed that the supportive (affiliative) management style dominated among managers, significantly influencing team cohesion and morale, although its effectiveness depended heavily on clear goal-setting and effective feedback mechanisms.

**Research limitations/implications**: The research underscores the need for managerial training to emphasize situational adaptability and the strategic combination of supportive and delegating leadership styles to enhance organizational effectiveness and economic performance.

**Practical implications:** The research provides practical guidelines for enhancing leadership development programs by promoting flexible use of management styles, which can improve employee engagement, decision-making efficiency, and ultimately drive better business performance and competitiveness.

**Social implications:** The research may positively impact society by promoting emotionally intelligent and adaptable leadership, fostering healthier workplace relationships, reducing conflict, and contributing to improved employee well-being and organizational culture.

**Originality/value:** The novelty of the paper lies in its practical application of Blanchard's situational leadership model to real business environments, offering empirical insights into managerial style correlations and adaptability, making it valuable for researchers, HR professionals, and organizational leaders aiming to improve leadership effectiveness.

**Category of the paper:** research paper.

# 1. Introduction

Managerial behavior remains a cornerstone of contemporary business management studies, reflecting the evolving nature of leadership theories and practices. Although existing literature extensively addresses various managerial competencies—ranging from technical and conceptual skills to emotional intelligence—there remains a significant gap concerning the dynamic adaptability of managerial styles in practice. The originality of this article lies in its practical exploration of leadership behaviors through the application of Kenneth Blanchard's diagnostic test, conducted among ten selected companies.

Unlike prior static examinations of leadership styles, this study uniquely focuses on situational adaptability and real-time decision-making processes within various business contexts. By employing a detailed questionnaire designed to provoke critical and analytical reflection among managers, this research captures nuanced insights into managerial behavior, revealing intricate relationships and potential trade-offs between distinct leadership styles such as instructional, supportive, delegating, and teaching. This practical, scenario-based methodological approach provides fresh empirical evidence of how specific managerial behaviors concretely impact team functionality, employee motivation, and organizational effectiveness.

Additionally, this research contributes original insights by statistically analyzing the correlations between different managerial styles, highlighting their potential synergies and conflicts. Understanding these interdependencies is particularly valuable for developing training programs and practical recommendations aimed at enhancing managerial effectiveness in diverse organizational environments. Thus, this article not only complements existing theoretical frameworks but significantly advances the understanding of effective managerial practice, emphasizing situational responsiveness and behavioral flexibility as key competencies for contemporary leaders.

## 2. Literature review on managerial roles, competencies, and leadership styles

The figure of the manager has long been a key topic in management theory. Józef Penc (1997) defines a manager as "a person employed in a leadership position, possessing broad knowledge and skills essential for directing people and managing organizations in conditions of uncertainty and constant change". The term itself, appearing interchangeably in Polish as "menadżer" or "menedżer". reflects the widespread adaptation of English managerial concepts into local contexts.

Effective management requires more than just theoretical education; it demands a specific set of skills and innate personality traits (Stoner, Wankel, 1992). Managerial competencies develop through a combination of academic preparation, professional experience, and constant exposure to practical challenges. Beyond theoretical knowledge, real-world practice allows managers to test their abilities, confront unexpected problems, and refine decision-making processes under dynamic conditions.

Managerial skills are traditionally categorized into three core groups: technical, interpersonal, and conceptual (Griffin, 1996). Some scholars also distinguish diagnostic and analytical skills as a complementary, yet crucial, set of competencies. Technical skills involve understanding the specialized knowledge and processes specific to the industry and organizational environment. Interpersonal skills emphasize the ability to communicate, collaborate, and motivate others effectively, ensuring smooth interpersonal relationships within and outside the organization (Mańczyk, 2001). Conceptual skills enable managers to perceive the organization holistically, fostering strategic thinking and abstract reasoning necessary for high-level decision-making. Diagnostic and analytical skills, on the other hand, support managers in systematically analyzing problems, forecasting outcomes, and selecting the most effective solutions.

The managerial hierarchy within organizations is another vital aspect of management science. Managers are often classified according to their level of responsibility: top-level managers (e.g., CEOs, directors), middle-level managers (e.g., department heads), and first-line managers (e.g., supervisors) (Penc, 1997). These roles vary significantly in terms of authority, scope of activities, and influence on organizational processes.

Additionally, managers can be distinguished by their functional areas, such as marketing, finance, operations, human resources, and administration (Koźmiński, 1999). Marketing managers, for instance, focus on customer acquisition and product promotion, whereas financial managers oversee resource allocation and investment planning. Operations managers optimize production systems, human resources managers handle staffing and personnel development, and administrative managers ensure the overall coordination of various business functions.

A crucial contribution to the understanding of managerial functions comes from Henry Mintzberg (2012), who identified ten essential managerial roles, grouped into three categories: interpersonal, informational, and decisional. Interpersonal roles include figurehead, leader, and liaison; informational roles encompass monitor, disseminator, and spokesperson; decisional roles involve entrepreneur, disturbance handler, resource allocator, and negotiator. Each role demands a specific set of behaviors and skills that enable managers to effectively perform their duties and adapt to the changing needs of the organization.

Classical management theory, particularly the work of Henri Fayol (1926 further emphasizes the functional aspects of management. Fayol divided organizational activities into six categories: technical, commercial, financial, security, accounting, and managerial. His focus on the managerial function led to the identification of five core tasks: planning, organizing, commanding, coordinating, and controlling. Planning involves setting objectives and developing action plans; organizing refers to the arrangement of resources; commanding is related to leading personnel; coordinating ensures the alignment of efforts; and controlling verifies the consistency of outcomes with initial plans.

Beyond technical and organizational competencies, modern leadership literature highlights the significance of emotional and social skills, particularly assertiveness. Assertiveness is often misunderstood as merely the ability to say "no"; however, it encompasses the broader capacity for open, honest, and respectful communication without infringing on the rights of others (Goleman, 1997; Bazan-Bulanda et al., 2020). Assertive individuals can express their needs, opinions, and feelings clearly while maintaining positive relationships and fostering trust within the organization.

Goleman (1997) and other scholars emphasize that assertiveness is not an inherent personality trait but a learnable skill, crucial for effective leadership. Assertive managers can establish healthy boundaries, manage interpersonal conflicts constructively, and maintain high levels of personal integrity and self-respect. They effectively balance their own rights and the rights of others, promoting a cooperative and empowering work environment.

Conversely, non-assertive behaviors - such as passive, aggressive, or pseudo-assertive communication - can undermine managerial effectiveness (Klos, 2017). Passive managers may fail to defend their positions, aggressive managers may damage workplace relationships, and pseudo-assertive managers may create confusion and distrust. Thus, developing true assertiveness skills is essential for sustainable leadership.

Internal dialogues and self-perceptions play a critical role in either supporting or sabotaging assertive behavior. Negative internal monologues filled with self-doubt or catastrophic thinking patterns can severely restrict a manager's ability to act assertively (Klos, 2017). Identifying and reframing such thought patterns is essential to building confidence and enhancing communication effectiveness.

Parween (2022) conducted an empirical study examining the influence of different leadership styles - democratic, autocratic, and instructional - on the productivity of academic staff in private universities in Vadodara, India. The results indicate that both democratic and instructional leadership styles have a positive impact on employee performance, whereas autocratic leadership was found to exert minimal influence. While the study offers valuable insights into leadership effectiveness within academic institutions, it does not address the situational flexibility of leaders or the extent to which leadership styles evolve in response to contextual demands. The present research seeks to build upon these findings by exploring the dynamic application of leadership styles across diverse managerial contexts, with particular emphasis on behavioral adaptation to changing circumstances (Parween, 2022).

Setiawan et al. (2021) highlight the significant impact that different leadership styles exert on employee productivity within organizational settings. Their comparative study reveals that autocratic leadership often correlates with reduced departmental efficiency, manifested in issues such as absenteeism and low morale. In contrast, egalitarian, transformational, and transactional leadership styles are shown to enhance employee engagement, motivation, and overall performance. These findings underscore the importance of adopting leadership styles that align with the psychological and operational needs of employees. However, the study offers a static perspective, focusing primarily on generalized outcomes rather than on the contextual application of leadership behaviors. It does not consider how leaders adjust their style in response to specific team dynamics or shifting organizational conditions (Setiawan et al.2021).

While classical theories such as Mintzberg's managerial roles and Fayol's functional approach provide a foundational understanding of managerial responsibilities, the present study operationalizes managerial behavior using Blanchard's situational leadership model. This model was selected due to its practical applicability and clear categorization of managerial responses across instructional, teaching, supportive, and delegating styles — which closely align with the interpersonal and adaptive competencies emphasized by Goleman (1997) and others. These constructs serve as the primary variables measured in our diagnostic questionnaire.

# 3. Research methodology and case description

Managers - regardless of the level at which they function - play a key role in building effective teams, shaping organizational culture, resolving conflicts, and motivating employees to act. In the face of dynamic market changes, increasing digitization and growing expectations of leaders, the question is increasingly being asked not so much "what a manager does", but "how he or she does it" - that is, what attitudes, styles and behaviors he or she displays in management practice. The behavior of managers has a direct impact on the functioning of entire organizations - it affects the atmosphere of work, the level of employee involvement, the efficiency of task performance and the readiness of teams to adapt in a volatile environment. Therefore, the analysis of these behaviors is becoming not only a subject of interest in management science, but also a real need for business practice. The identification of dominant management styles can contribute to a better understanding of the mechanisms of people management and the formulation of recommendations for improving managerial competence.

The study aims to highlight an in-depth analysis of managers' behavior in the context of managing human teams in a company. In particular, it focuses on the identification of management styles, their determinants and potential consequences for the functioning of the team and the organization as a whole. To this end, a research methodology was developed based on a proven diagnostic tool - Kenneth Blanchard test - known for its practical application in identifying management styles. The study was conducted in a group of 10 selected companies. The primary research tool was a specially developed diagnostic questionnaire, consisting of 20 questions. The questionnaire was constructed on the basis of the Blanchard test, which is used to assess managerial behavior and identify management styles. The survey was situational in nature - the questions presented a specific situation that could occur in the daily operation of a company. The survey participant (manager/leader) was asked to select one of four possible answers (labeled A-D), which reflected different styles and strategies for responding to a given situation. An example question was: "Your team has recorded excellent job performance over the past two years. However, due to factors beyond their control, your employees have recently experienced three major failures. Their morale and work performance have dropped dramatically and your boss is concerned about this. How would you behave in a team meeting?" Questions of this type were designed to stimulate critical and analytical thinking, as well as reflection on one's own leadership style. Other questions focused on interpersonal relations within the team, the level of trust, the ability to motivate and delegate tasks, and openness to innovation and new ideas.

Each response represented a specific management style: instructional, supportive, delegating or teaching. The instructive style is based on precise goal setting and detailed instruction of the employee. In this case, the leader assumes the role of teacher and organizer, who not only plans and sets tasks, but also actively supervises their implementation. This requires a high level of commitment, patience and developed communication skills from the supervisor, especially in terms of listening and giving clear instructions. The teaching style, combines leadership with active support of the employee. The leader in this case provides the necessary information and instructions, but at the same time leaves room for independent decision-making. The key task of the manager is to build the employee's self-confidence, encourage initiative and allow him to choose the best solution to the problem in his opinion. The supportive style, also known as affiliative, implies a more collaborative relationship between the leader and the team member. The manager focuses on building trust, open communication and sharing his experience. By actively listening, giving praise and helping to overcome difficulties, the leader assists the employee in developing autonomy and decision-making. The delegating style, often referred to as autocratic-visionary, is characterized by a high level of employee autonomy. The leader consciously delegates responsibility for daily decisions to team members, supporting their independence and initiative. He or she encourages new challenges and creates conditions for independent action, withdrawing from direct influence on decision-making, but remaining a mentor and visionary providing direction.

After collecting the responses, the data was entered into an analysis table, where each response was assigned a corresponding management style. The analysis sheet consisted of four columns corresponding to different management styles, and the sum of the scores in each column indicated the dominant style of a given manager. Thus, the questionnaire made it possible not only to diagnose the current management style, but also to identify trends and possible disparities in the use of certain management strategies. This approach allowed an in-depth qualitative and quantitative analysis of managerial behavior in the surveyed companies.

## 4. Research results

A summary of the survey results for all 10 companies, corresponding to the 20 survey questions asked, is presented in pie charts from 1 to 12 and in Table number 1 which shows the system for collecting responses and transforming them into a preferred management system. The table is arranged according to management styles (instructive style, teaching style, supportive style, delegating style) in an identical manner for each survey participant as shown in Table 1.

In the charts showing the results of the Blanchard test for each company, the management styles dominant in that company have been bolded in the legend. This makes it possible to quickly identify the preferred management style for each leader, facilitating comparison between companies and analysis of the distribution of styles across the research sample.

**Table 1.**
*Preferred management style according to Blanchard test for company number 1*

| Survey question number | Types of management styles | | | |
|:---:|:---:|:---:|:---:|:---:|
| | Instructional style | Teaching style | Supportive style | Delegating style |
| 1 | A | B | C | D |
| 2 | A | B | C | D |
| 3 | A | B | C | D |
| 4 | A | B | C | D |
| 5 | A | B | C | D |
| 6 | A | B | C | D |
| 7 | A | B | C | D |
| 8 | A | B | C | D |
| 9 | A | B | C | D |
| 10 | A | B | C | D |
| 11 | A | B | C | D |
| 12 | A | B | C | D |
| 13 | A | B | C | D |
| 14 | A | B | C | D |
| 15 | A | B | C | D |
| 16 | A | B | C | D |
| 17 | A | B | C | D |
| 18 | A | B | C | D |
| 19 | A | B | C | D |
| 20 | A | B | C | D |
| Total points | 5 | 5 | 6 | 4 |

Analysis of the collected data showed the following breakdown of the dominant management styles in the sample:

- Supportive (affiliative) style:
  o The dominant style in 6 of the 10 companies surveyed, accounting for 60% of the sample analyzed.
  o This style received the highest total number of indications in all participant responses.
- Delegating style:
  o The delegating style emerged as the dominant style in 3 companies, accounting for 30% of the sample.
  o The number of indications for the delegating style was in second place compared to the other styles.
- Teaching style:
  o The teaching style was identified as dominant in only 1 company (10% of the sample).
  o This style occurred with low frequency compared to the other two styles.
- Instructional style:
  o None of the managers surveyed identified instructional style as dominant, resulting in a score of 0% in this category.

The summary of the results in both numerical and percentage form makes it possible to clearly state that the most frequently emerged, dominant management style was a supportive attitude, which is confirmed by both individual assessments and the combined analysis of all respondents.

In addition, the results are visualized in the form of graphs shown in Figures 1-10, where the style that was marked as dominant in a given company is bolded in the legend of each graph. This makes it possible to quickly compare the results between different companies and get a clear picture of the distribution of management styles in the surveyed group.

The results are visualized in the form of graphs shown in Figures 1-10, where the style that is marked as dominant in a given company is bolded in the legend of each graph. This makes it possible to quickly compare the results between different companies and get a clear picture of the distribution of management styles in the studied group.

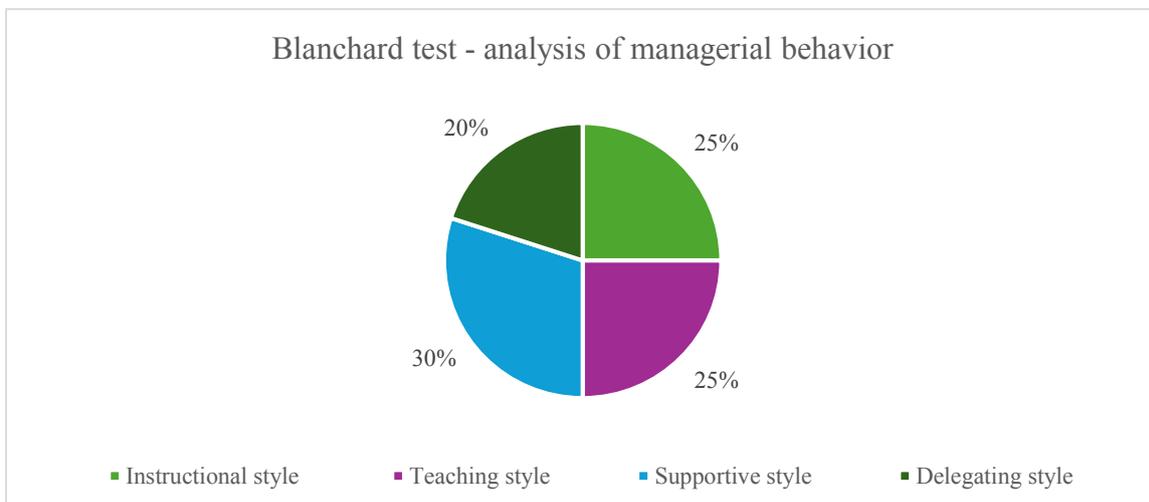

**Figure 1.** Summary of answers of the Leader of the enterprise number 1 according to the Blanchard test.

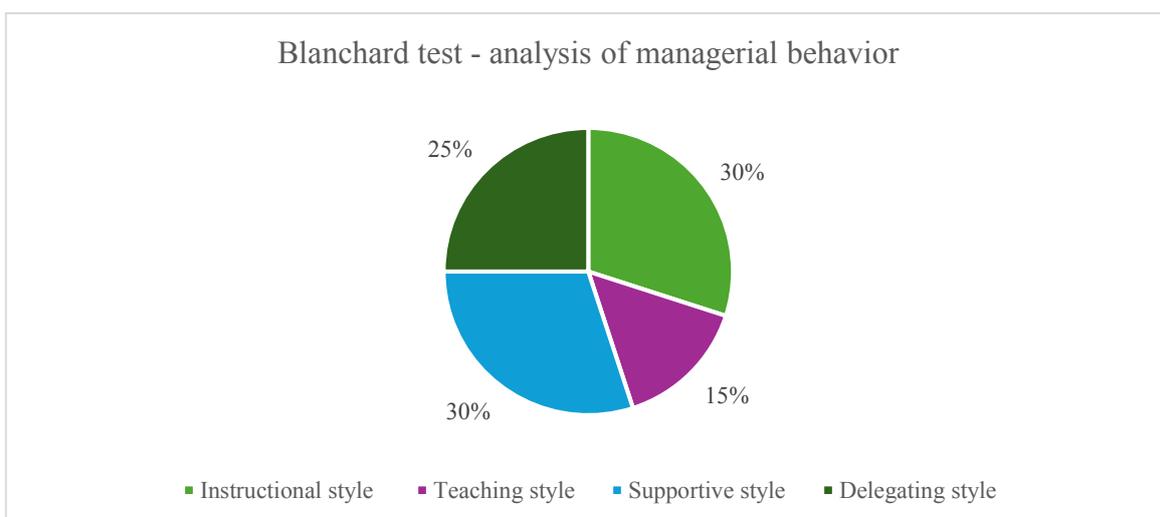

**Figure 2.** Summary of answers of the Leader of the enterprise number 2 according to the Blanchard test.

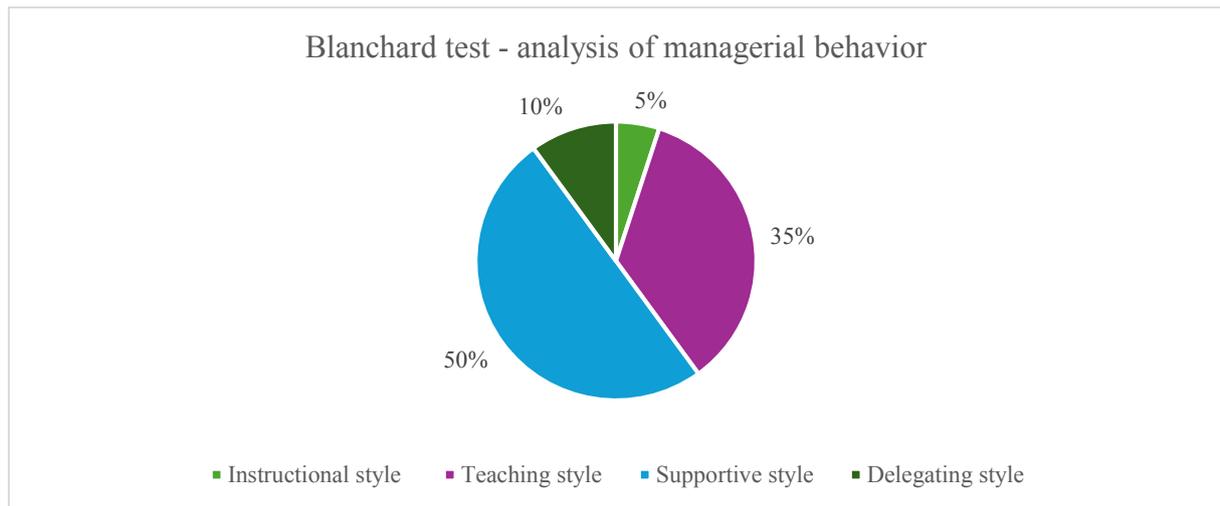

**Figure 3.** Summary of answers of the Leader of the enterprise number 3 according to the Blanchard test.

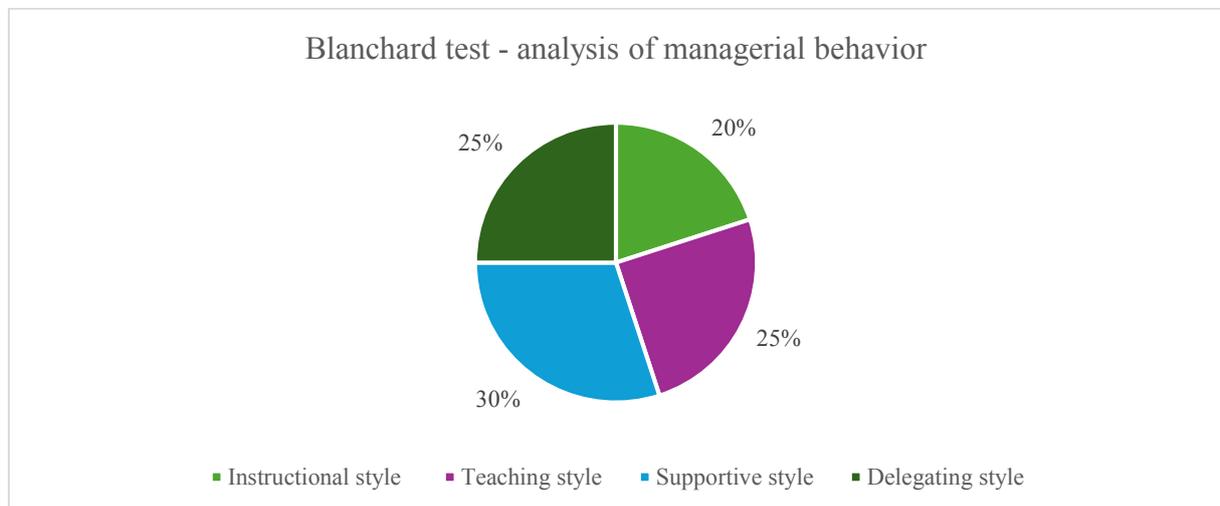

**Figure 4.** Summary of answers of the Leader of the enterprise number 4 according to the Blanchard test.

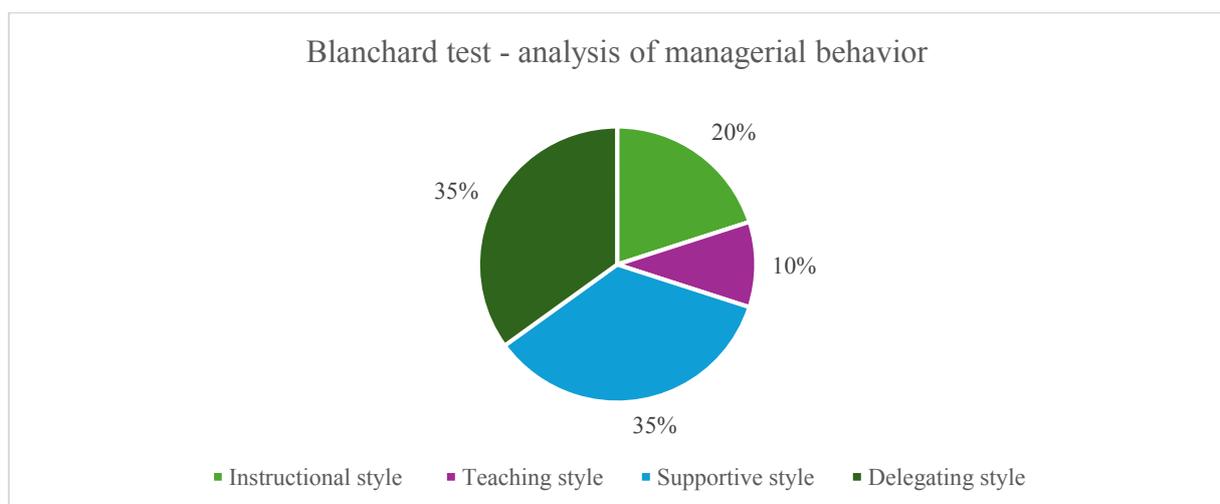

**Figure 5.** Summary of answers of the Leader of the enterprise number 5 according to the Blanchard test.

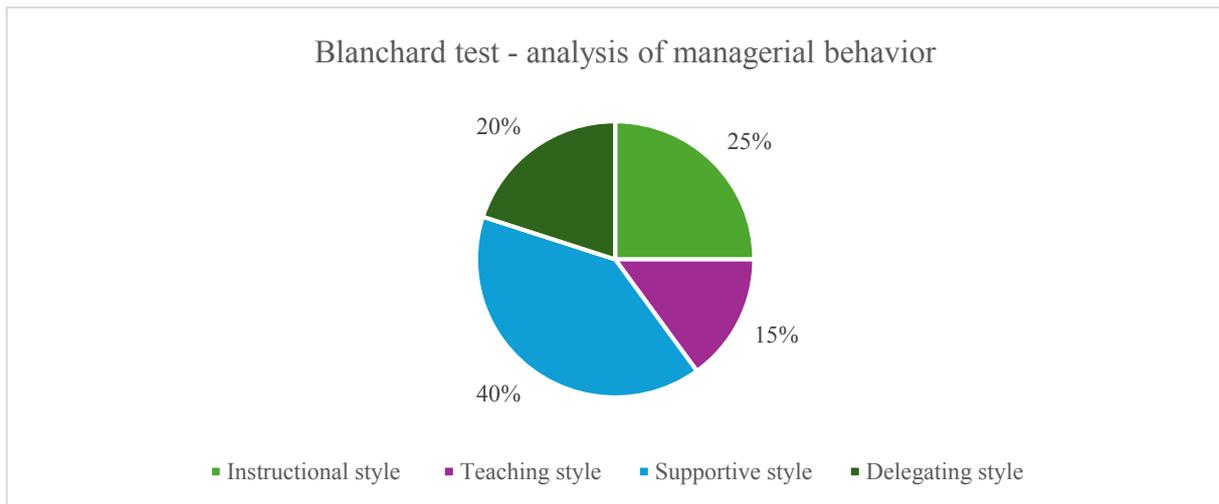

**Figure 6.** Summary of answers of the Leader of the enterprise number 6 according to the Blanchard test.

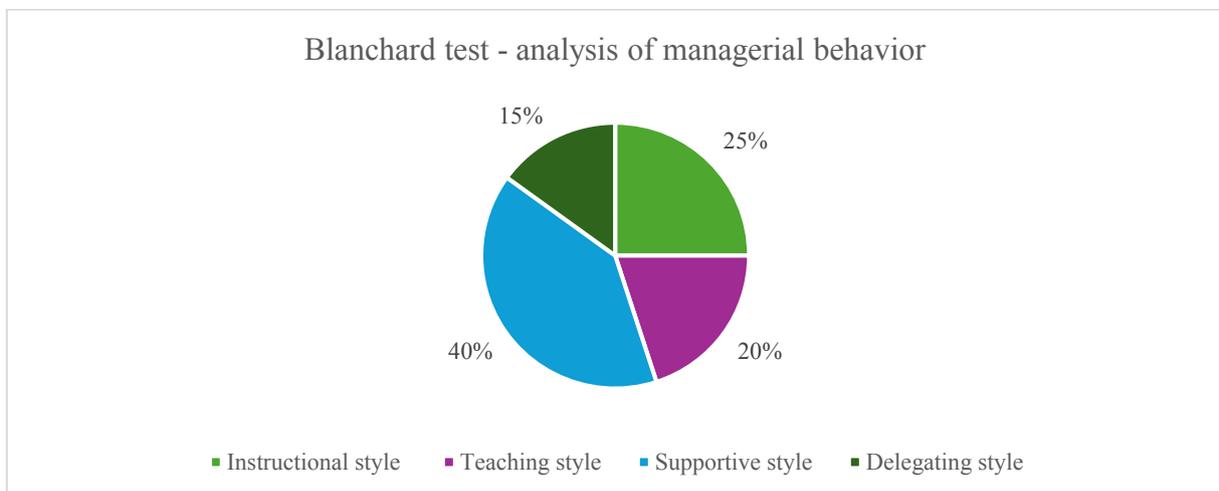

**Figure 7.** Summary of answers of the Leader of the enterprise number 7 according to the Blanchard test.

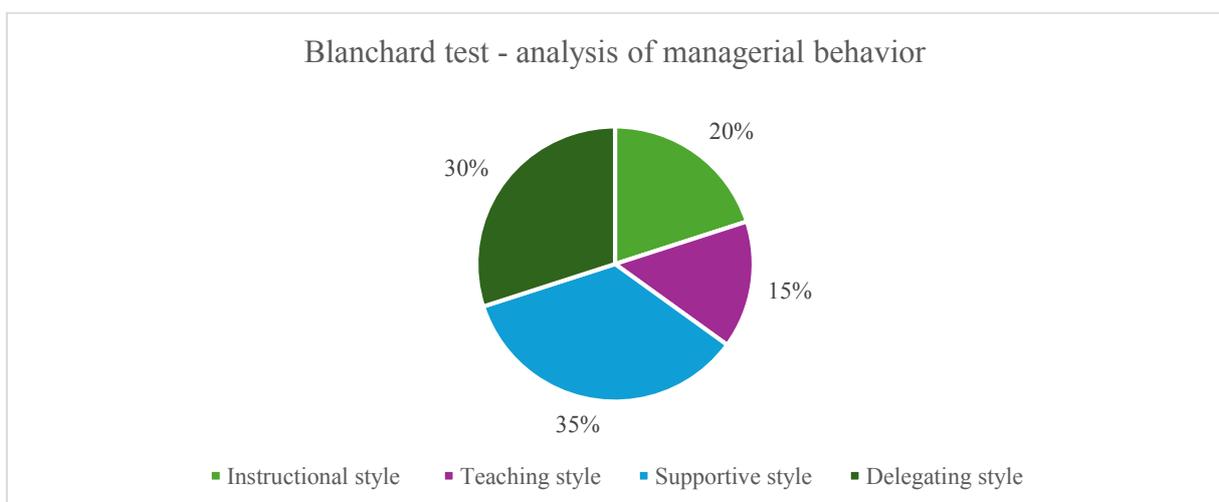

**Figure 8.** Summary of answers of the Leader of the enterprise number 8 according to the Blanchard test.

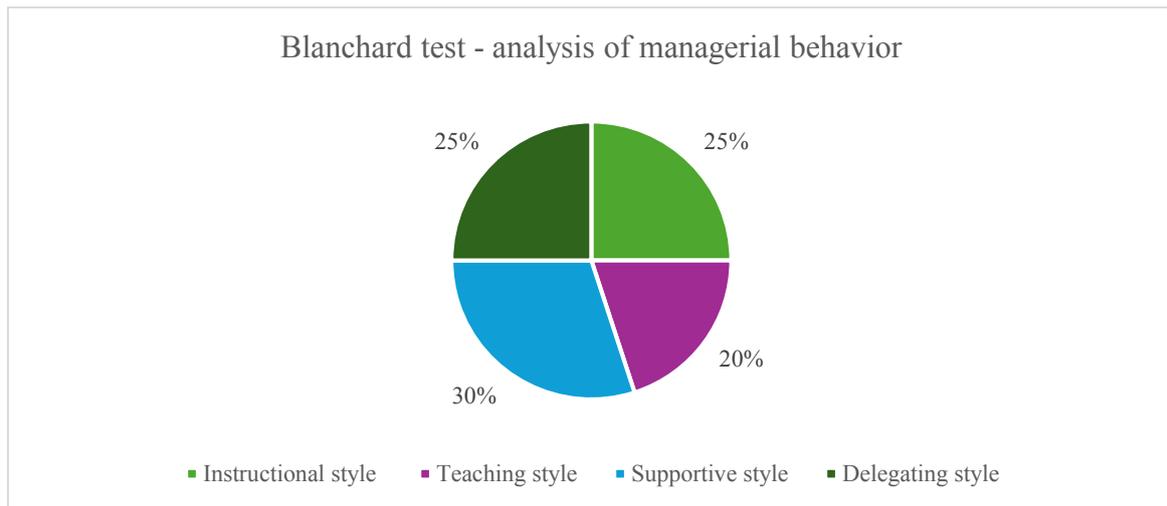

**Figure 9.** Summary of answers of the Leader of the enterprise number 9 according to the Blanchard test.

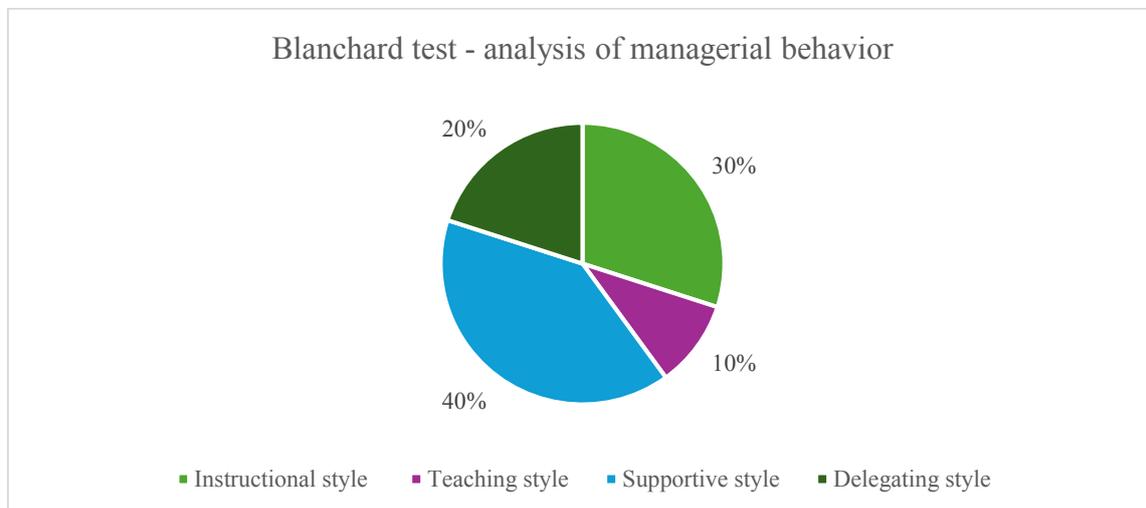

**Figure 10.** Summary of answers of the Leader of the enterprise number 10 according to the Blanchard test.

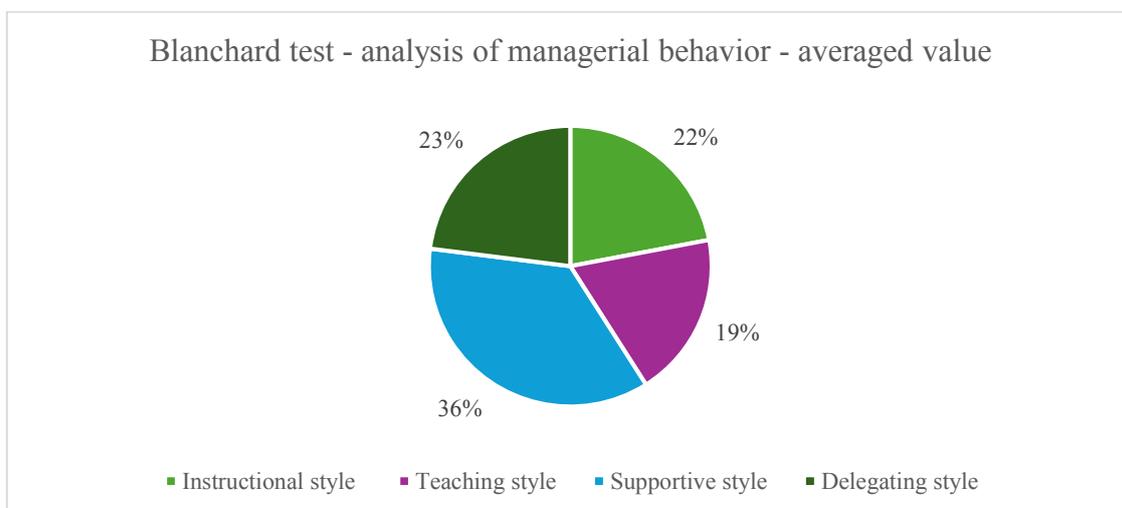

**Figure 11.** Summary of Leader's responses for all companies according to the Blanchard test.

In order to determine the statistical relationships between the different styles, a correlation analysis was conducted based on the collected scores assigned to each management style. The results of the correlation analysis are shown in Figure 12 (correlation matrix).

The results presented in the correlation matrix document the level of interdependence between management styles, which is an additional dimension of the analysis of the data obtained. All indicators were calculated based on the self-assessment of managers completing the diagnostic questionnaire.

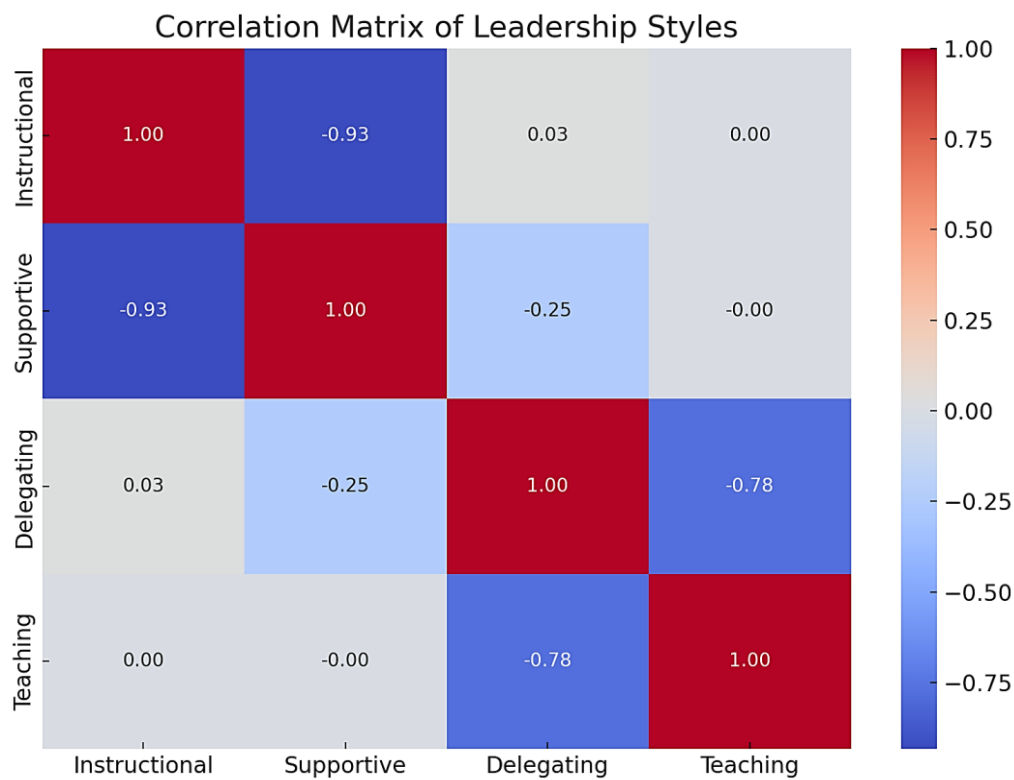

**Figure 12.** Correlation Matrix of Management Styles according to the conducted Blanchard test.

An analysis of correlations between management styles revealed significant relationships that have practical implications for the development of managerial competence. The strong negative correlation between instructional and supportive styles (r = -0.93) indicates that the more often a leader uses a style based on control and issuing orders, the less often he or she exhibits attitudes focused on relationships and emotional support of the team - and vice versa. The instructional and supportive styles thus represent opposing approaches to management.

In contrast, the lack of a significant correlation between instructional and delegating styles (r ≈ 0.03) suggests that the two styles function independently - managers can be simultaneously instructional in some situations and delegating in others, confirming the flexibility of their management approach. An interesting relationship is also revealed by the strong negative correlation between teaching and delegating styles (r = -0.78), which means that leaders who support employees' development and decisions (teaching) are less likely to completely delegate autonomy to employees. This may be due to the need to maintain close contact, mentoring and ongoing support rather than completely surrendering control. On the other hand, the lack of a significant correlation between teaching and supportive styles suggests that although the two styles have similar assumptions (development, support), in practice they are used independently - probably due to differences in leaders' competencies or preferences.

For example, if a leader in Company 5 exhibits high in supportive style and low in instructional style, this may indicate his effectiveness in managing a team with a high need for emotional security. However, such a leader may have difficulty enforcing results in crisis situations or low team motivation.

From the point of view of management practice, the development of managerial competence should be based on flexibility - the best leaders are able to consciously switch between different styles depending on the situation, the level of employee development and organizational goals. Therefore, training programs for managers should include elements of training in recognizing the situation, assessing the needs of the team and consciously adjusting the leadership style.

The best management style is one that strengthens employees' abilities and confidence. In the long run, it fosters the development of the entire team. A leader who focuses on the strengths and weaknesses of his people helps them discover their own aspirations and values, supports their development and gives them a sense of fit with the organization. Such a leader not only boosts motivation, but also encourages employees to continuously improve and share responsibility for results.

## 5. Discussion

The results of this study provide a deeper understanding of how managerial behavior—particularly leadership style—shapes team dynamics, employee motivation, and organizational performance. By applying Kenneth Blanchard's diagnostic tool in ten diverse companies, this research offers empirical confirmation that the supportive (affiliative) style is currently dominant in many business environments. This finding aligns with previous literature emphasizing the importance of interpersonal skills, emotional intelligence, and trust-based relationships in modern leadership (Goleman, 1997; Bazan-Bulanda et al., 2020).

One of the main contributions of this study is the confirmation that leadership flexibility is a critical determinant of effectiveness. The analysis revealed that certain styles, such as instructional and supportive, tend to function as opposites ($r = -0.93$), while others, such as delegating and instructional, may coexist depending on context ($r \approx 0.03$). This reinforces the theoretical perspective of situational leadership (Blanchard), in which leaders must diagnose specific team needs and adapt their style accordingly.

Compared to prior studies (Parween, 2022; Setiawan et al., 2021), which largely present leadership styles as fixed or static traits, this study extends the existing knowledge by highlighting behavioral adaptability. For example, while Setiawan et al. demonstrated that autocratic leadership reduces employee engagement, our findings show that no single style is universally effective or ineffective—the key lies in matching the style to the maturity of the team and task complexity.

This research also confirms the theoretical proposition that over-reliance on one leadership style—even a people-centered one like the supportive style—can lead to unintended negative outcomes if not supplemented with other managerial tools such as clear goal-setting, constructive feedback, and structured delegation. While supportive leadership builds morale, a lack of direction may impair performance. These findings echo Goleman's assertion that empathy must be balanced with assertiveness for leadership to be sustainable and effective.

Furthermore, the lack of strong correlation between teaching and supportive styles suggests that, in practice, leaders may favor different tools depending on their comfort zone or contextual familiarity, not necessarily based on logical complementarity. This highlights the need for further leadership training aimed at improving self-awareness and versatility.

In terms of theoretical advancement, this study contributes to the understanding of style interdependence and suggests a new dimension to situational leadership: not just adapting to employee maturity, but consciously integrating or switching styles across tasks and time. This complements Mintzberg's idea of multifaceted managerial roles, where a leader must simultaneously serve as a strategist, communicator, motivator, and resource allocator.

From a practical standpoint, our findings suggest that organizations should not only train leaders in individual styles, but also in style integration and sequencing. Diagnostic tools like the Blanchard test should be incorporated into leadership development programs to help managers identify their dominant tendencies and potential blind spots.

Despite these contributions, several limitations should be noted. The sample size (10 companies) and reliance on self-assessment surveys limit the generalizability of findings. Additionally, as the study captured a cross-sectional perspective, it does not account for how leadership styles evolve over time or during organizational crises.

Prior studies emphasized the effectiveness of democratic, transformational, or supportive styles in isolation. This study adds depth by exploring how combinations and transitions between styles matter, offering a more dynamic and actionable view of leadership behavior.

Future research should explore longitudinal changes in managerial behavior, including how leaders adapt over time, under stress, or through feedback. Comparative studies across industries and cultures would also enrich our understanding of how context shapes effective leadership.

## 6. Conclusions and future research implications

The conducted study provides valuable insights into the behavioral preferences and leadership styles of managers across ten companies, based on the diagnostic framework of the Blanchard test. The dominant presence of the supportive (affiliative) management style across the sample indicates a significant emphasis on interpersonal relations, team cohesion, and employee well-being. This aligns with contemporary leadership literature, which emphasizes emotional intelligence, assertiveness, and trust-building as critical elements of effective management (Goleman, 1997; Bazan-Bulanda et al., 2020).

Our findings reinforce the theoretical framework presented in earlier studies (Parween, 2022; Setiawan et al., 2021), which show that leadership styles oriented toward collaboration and psychological safety tend to foster higher levels of employee motivation and organizational harmony. The supportive style identified in this research corresponds closely with democratic and transformational leadership, previously linked with improved productivity and engagement. However, the study expands on these findings by showing that while affiliative leadership builds strong bonds, it must be complemented by clear goal setting and feedback mechanisms to avoid inefficiencies and performance declines.

The analysis of style correlations adds an additional layer of understanding, revealing that some leadership styles (e.g., instructional and supportive) function in opposition, while others (e.g., instructional and delegating) may coexist flexibly depending on the context. This underscores the importance of situational leadership flexibility. Our findings suggest that effective leaders adapt their styles fluidly, depending on employee maturity, task complexity, and environmental volatility.

Practical implications of this research are considerable for both managerial training and organizational development:

- Leadership development programs should prioritize teaching managers to diagnose situational variables and consciously adjust their behavior.
- Supportive leadership should be complemented by assertive feedback skills and vision-based delegation, especially in fast-changing or performance-oriented environments.
- Organizations may consider integrating diagnostic tools such as the Blanchard test into their HR practices to foster self-awareness and style adaptability among leaders.
- Creating mixed-style leadership environments (e.g., combining supportive and delegating approaches) may help balance emotional engagement with performance outcomes.

Despite its contributions, this study has several limitations:

- The sample was relatively small (10 companies), limiting the generalizability of findings across industries and cultures.
- The data was derived from self-assessment questionnaires, which may be influenced by social desirability bias or self-perception distortions.
- The study presents a cross-sectional snapshot of leadership behavior, without accounting for changes over time or under different organizational conditions.

Future research should consider longitudinal approaches to better understand how leadership styles evolve in response to external shocks, organizational transitions, or personal development. Expanding the research across different industries, cultural contexts, and company sizes would also offer broader applicability. Moreover, future studies could examine how leadership style combinations (hybrid styles) impact specific outcomes such as innovation rates, staff retention, or resilience in times of crisis. Given the growing importance of digital transformation and remote leadership, additional focus on how management styles translate into virtual team settings would be highly beneficial.